\documentclass[preprint,nofootinbib]{revtex4}

\usepackage{graphicx}
\usepackage{epsfig}

\begin{document}

\title{Scalar field black holes}

\author{S. Carneiro$^1$\footnote{In sabbatical from Instituto de F\'{\i}sica, Universidade Federal da Bahia, 40210-340, Salvador, Bahia, Brazil.} and J. C. Fabris$^{2,3}$}

\affiliation{$^1$Instituto de F\'{\i}sica Gleb Wataghin - UNICAMP, 13083-970, Campinas, SP, Brazil\\$^2$Departamento de F\'{\i}sica - UFES, 29075-910, Vit\'oria, ES, Brazil\\$^3$National Research Nuclear University MEPhI, Moscow, Russia}

\date{\today}

\begin{abstract}
With a suitable decomposition of its energy-momentum tensor into pressureless matter and a vacuum type term, we investigate the spherical gravitational collapse of a minimally coupled, self-interacting scalar field, showing that it collapses to a singularity. The formed blackhole has a mass $M \sim 1/m$ (in Planck units), where $m$ is the mass of the scalar field. If the latter has the axion mass, $m \sim 10^{-5}$ eV, the former has a mass $M \sim 10^{-5} M_{\odot}$.
\end{abstract}

\maketitle

The interface between cosmology and blackhole physics has been always fertile, leading to theoretical insights and mathematical analogies that go from singularity theorems to Hawking radiation and holography \cite{gibbons1,gibbons2,gibbons3,gibbons4,dain}. This interface is a counterpart of another rich interplay between process occurring in the primordial universe and those characteristic of dense nuclear media like relativistic stars. In the cosmological realm, a relatively recent approach to the dark sector has been based on unified descriptions where both the dark energy and the dark matter are described by a single fluid, formed for example by a generalised Chaplygin gas \cite{ggc1,ggc2,ggc3,ggc4,ggc5}, or by a non-adiabatic scalar field \cite{scalar1,scalar2,scalar3,sigma1,sigma2}. In this context, it is natural to ask whether such unified fluids can also allow for star-like and blackhole solutions \cite{bertolami1,bertolami2}. Scalar field blackholes have been investigated in de Sitter and anti-de Sitter backgrounds, in both conformal and massive cases  \cite{zanelli1}. Blackhole and wormhole solutions have been found with phantom fields as well \cite{bronnikov}. From an astrophysical viewpoint, these solutions are particularly interesting in the case of an axion field \cite{zeldovich,kolb}, the only stable scalar predicted by the standard model of particles, and one of the natural candidates for the unobserved dark matter particle. Axionic blackholes are also interesting because of the possible signature the accretion of axions can leave on future gravitational waves observations \cite{GW}. Accretion of scalar fields into blackholes was studied, for example, in \cite{alberto}.

A minimally coupled scalar field has energy-momentum tensor
\begin{equation}
T^{\mu\nu} = \partial^{\mu}\phi \, \partial^{\nu}\phi + \left( V - \frac{1}{2} \partial_{\alpha}\phi \, \partial^{\alpha}\phi \right) g^{\mu\nu},
\end{equation}
where $V(\phi)$ is the self-interaction potential. We can formally decompose it into pressureless and ``vacuum" components
\begin{eqnarray}\label{matter}
T_m^{\mu\nu} &=& \partial^{\mu}\phi \, \partial^{\nu}\phi, \\ \label{lambda}
T_{\Lambda}^{\mu\nu} = \Lambda  g^{\mu\nu} &=& \left( V - \frac{1}{2} \partial_{\alpha}\phi \, \partial^{\alpha}\phi \right) g^{\mu\nu}.
\end{eqnarray}
The ``matter" density is defined as $\rho = T_{m\mu}^{\mu}$, and the scalar field $4$-velocity by
\begin{equation} \label{velocidade}
u^{\mu} = \frac{\partial^{\mu} \phi}{\sqrt{\partial^{\alpha}\phi \, \partial_{\alpha} \phi}}.
\end{equation}
The vacuum density $\rho_{\Lambda}=\Lambda$ is a covariant scalar, and its equation of state, $p_{\Lambda}=-\rho_{\Lambda}$, is the same for any observer.
Using the above decomposition, it was shown elsewhere that a scalar field can accomplish for both dark components observed in the cosmic fluid \cite{sigma1,sigma2}. The $\Lambda$ component does not cluster at linear order and is responsible for the expansion acceleration. The pressureless one, on the other hand, can be identified with observed clustering matter. With an appropriate choice of the scalar field potential, we have the same phenomenology of non-adiabatic generalised Chaplygin gases, which includes the standard $\Lambda$CDM particular case. Let us show here, using the same decomposition, that a massive scalar field can perform a spherical gravitational collapse, leading to the formation of blackholes.

In comoving coordinates, the metric inside the fluid region of a spherically symmetrical spacetime can be written as
\begin{equation} \label{metric}
ds^2 = e^{\nu} d\tau^2 - e^{\lambda} dR^2 - e^{\mu} (d\theta^2 + \sin^2 \theta d\phi^2), 
\end{equation}
where $\nu$, $\lambda$ and $\mu$ are generic functions of time and of the radial coordinate $R$. For a substratum formed by a perfect fluid like above, with energy-momentum tensor
\begin{equation}
T^{\mu\nu} = \rho u^{\mu} u^{\nu} +
\Lambda  g^{\mu\nu},
\end{equation}
the independent Einstein equations are given by \cite{Landau}
\begin{eqnarray}\label{L2}
\Lambda &=& e^{-\nu} \left( \ddot{\mu} - \frac{\dot{\mu}\dot{\nu}}{2} + \frac{3\dot{\mu}^2}{4} \right) - \frac{e^{-\lambda}}{2} \left( \frac{\mu'^2}{2} + \mu' \nu' \right) +
e^{-\mu},\\ \label{L4}
\Lambda + \rho &=& \frac{e^{-\nu}}{2} \left( \dot{\mu}\dot{\lambda} + \frac{\dot{\mu}^2}{2} \right) - e^{-\lambda} \left( \mu'' + \frac{3\mu'^2}{4} - \frac{\mu' \lambda'}{2} \right) + e^{-\mu},\\ \label{L5}
0 &=& 2 \dot{\mu}' + \dot{\mu} \mu' - \dot{\lambda} \mu' - \nu' \dot{\mu},
\end{eqnarray}
where a dot means derivative w.r.t. time and a prime means derivative w.r.t. $R$. On the other hand, the Bianchi identities $T^{\mu \nu}_{;\nu} = 0$ lead to
\begin{eqnarray}\label{L6a}
\nu' = \frac{2\Lambda'}{\rho},\\ \label{L6b}
\dot{\rho} + \Theta \rho = -\dot{\Lambda},
\end{eqnarray}
where
\begin{equation}\label{continuidade}
\Theta = \frac{\dot{\lambda}}{2} + \dot{\mu}.
\end{equation}
For comoving observers the $\Lambda$ component remains homogeneous, that is, $\Lambda = \Lambda (\tau)$. Hence, from (\ref{L6a}) we see that $\nu = \nu(\tau)$, and this allows the choice of synchronous observers, for which $\nu = 0$. By also defining $r^2 = e^{\mu}$, we have from (\ref{L5})
\begin{equation} \label{e^lambda}
e^{\lambda} = \frac{r'^2}{1 + f(R)},
\end{equation}
where $f(R)$ is an arbitrary function of $R$. In this way, it is straightforward to obtain, from (\ref{L2}) and (\ref{L4}), the equations
\begin{eqnarray}\label{p}
\Lambda r^2 &=& 2 r \ddot{r} + \dot{r}^2 - f(R),\\ \label{rho}
\rho r^2 &=& -2 r \ddot{r} - \frac{rf'}{r'} + \frac{2r\dot{r}\dot{r}'}{r'}.
\end{eqnarray}
From (\ref{continuidade}) we also have
\begin{equation}\label{Theta}
\Theta = \frac{1}{r'r^2} \frac{d(r'r^2)}{d\tau}.
\end{equation}

Consider a scalar field that is initially homogeneous and infinitely diluted, and let us assume that the $\Lambda$ component is positive and remains constant along the collapse. In this case, equations (\ref{p}) and (\ref{rho}) are identically satisfied if we take
\begin{eqnarray}\label{solution1}
\dot{r}^2 &=& \frac{F(R)}{r} + f(R) + \frac{\Lambda r^2}{3},
\end{eqnarray}
\begin{equation} \label{solution2}
\rho = \frac{F'(R)}{r'r^2},
\end{equation}
where $F(R)$ is another arbitrary function of $R$, with $F(0) = 0$ in order to have $r(0,t) = 0$. With this it is easy to check that (\ref{L6b}) (with $\Theta$ given by (\ref{Theta})) is also verified. Solutions (\ref{solution1}) and (\ref{solution2}) characterise the collapse of the pressureless component into the origin. The two arbitrary functions $F(R)$ and $f(R)$ determine the density and radial velocity profiles.
When the event horizon is formed, the interior metric (\ref{metric}) (with $\nu = 0$) can be matched with an exterior Schwarzschild metric. The junction conditions are given by \cite{julio}\footnote{In \cite{julio} metric (\ref{metric}) is matched with an exterior Schwarzschild-de Sitter metric, whereas in the presente case the resulting spacetime is asymptotically flat.}
\begin{eqnarray}\label{J1}
0 &=& \left( X \dot{Y}' - \dot{X} Y' \right)_{R \rightarrow \infty},\\ \label{J2}
r_g &=& \left[ Y \left(1 + \dot{Y}^2 - \frac{Y'^2}{X^2} \right) \right]_{R\rightarrow \infty},
\end{eqnarray}
where $r_g$ is the gravitational radius, $X^2 = e^{\lambda}$ and $Y^2 = e^{\mu} = r^2$. The former of these conditions is identically satisfied by (\ref{e^lambda}). The latter leads to $\dot{r}^2 |_{r_g} = 1 + f(\infty)$. Hence we have, from (\ref{solution1}),
\begin{equation}\label{r_g}
r_g \left( 1 - \frac{\Lambda}{3} r_g^2 \right) = F(\infty).
\end{equation}
For $F(\infty) = 0$, we have $r_g = \sqrt{3/\Lambda}$, corresponding to a de Sitter core matched with an asymptotically flat spacetime \cite{Lemos}. For $F(\infty)\neq0$, the solution of (\ref{r_g}) is unique only for
\begin{equation}\label{F}
F(\infty) = \frac{2}{3\sqrt{\Lambda}},
\end{equation}
given by $r_g = 1/\sqrt{\Lambda}$. The collapsed mass is
\begin{equation}\label{mass}
M = 4\pi r_g = \frac{4\pi}{\sqrt{\Lambda}} =  4\pi F(\infty) + \frac{4\pi}{3} r_g^3 \Lambda.
\end{equation}
The last equality has an evident interpretation if, by using (\ref{solution2}), we note that
\begin{equation}
\int_0^{\infty} \rho r^2 r' dR = F(\infty).
\end{equation}

In the present configuration, in comoving coordinates the scalar field is homogeneous (see (\ref{velocidade})) . Therefore, from (\ref{solution2}) we see that $r'r^2 = F' h(\tau)$, where $h(\tau)$ is a function to be determined. Taking for $r$ the separable solution $r = g(R) [h(\tau)]^{1/3}$, we obtain $F' = g' g^2$. From (\ref{solution1}) we also have, taking the particular case $f(R) = 0$,
\begin{equation}
3Fg^{-3} = \frac{\dot{h}^2}{3h} - \Lambda h = k,
\end{equation}
where $k$ is a separation constant. From these equations it is easy to see that $k = 1$, $g = (3F)^{1/3}$, and
\begin{equation}
\Lambda h = \sinh^2 \left[ \frac{\sqrt{3\Lambda}}{2} \left( \tau - \tau_0 \right) \right],
\end{equation}
where $\tau_0$ is an integration constant.
Finally, we have the solution
\begin{equation}\label{r}
r(\tau,R) = \left[ \frac{3F(R)}{\Lambda} \right]^{1/3} \sinh^{2/3} \left[ \frac{\sqrt{3\Lambda}}{2} \left( \tau - \tau_0 \right) \right],
\end{equation}
while for the matter density we obtain
\begin{equation}\label{densidade}
\rho = \frac{\Lambda}{\sinh^2 \left[ \frac{\sqrt{3\Lambda}}{2} \left(\tau - \tau_0\right)\right]}.
\end{equation}
This uniform density goes to zero for $\tau \rightarrow -\infty$, and diverges for $\tau \rightarrow \tau_0$.
With solution (\ref{r}), metric (\ref{metric}) acquires the form\footnote{The choice $F = R^3$ (which does not obey the boundary condition (\ref{F})) corresponds to a spatially flat FLRW space-time filled with pressureless matter and a cosmological constant.}
\begin{equation}
ds^2 = d\tau^2 - a^2(\tau) \left[ \frac{F'^2}{9F^{4/3}} dR^2 + F^{2/3} (d\theta^2 + \sin^2 \theta d\phi^2) \right],
\end{equation}
where
\begin{equation}
a(\tau) = \left( \frac{3}{\Lambda} \right)^{1/3} \sinh^{2/3} \left[ \frac{\sqrt{3\Lambda}}{2} \left( \tau - \tau_0 \right) \right].
\end{equation}
It is a generalisation of the parabolic LTB metric (with uniform density) in presence of a vacuum term \cite{julio}. At any finite time, the scalar field distribution has a finite physical radius $r(\tau,\infty) = a(\tau) F^{1/3}(\infty)$. Its surface crosses the horizon at a time given by $r(\tilde{\tau},\infty) = r_g$. Using (\ref{r}), (\ref{mass}) and (\ref{F}), we obtain the collapse time
\begin{equation}
\tau_0 - \tilde{\tau} = \frac{2}{\sqrt{3\Lambda}} \sinh^{-1} \left( \frac{\sqrt{2}}{2} \right).
\end{equation}

Let us give the conditions on the scalar field potential in order to have a constant $\Lambda$. From (\ref{lambda}) it must satisfy
\begin{equation}\label{condition'}
V = \Lambda + \frac{\dot{\phi}^2}{2} - \frac{\phi'^2}{2} e^{-\lambda},
\end{equation}
and from (\ref{matter}) we have 
\begin{equation}\label{rho'}
\rho = 2(V-\Lambda).
\end{equation}
By using (\ref{densidade}) and (\ref{condition'}), with $\phi' = 0$, into (\ref{rho'}) we obtain, after integration in $\tau$,
\begin{equation}
\rho = \Lambda \sinh^2 \left( \frac{\sqrt{3}\phi}{2} \right),
\end{equation}
and the potential\footnote{It is the same potential adopted for a unified description of the $\Lambda$CDM cosmological dark sector by means of a single scalar field \cite{sigma2}.}
\begin{equation}\label{potential}
V(\phi) = \Lambda + \frac{\Lambda}{2} \sinh^2 \left( \frac{\sqrt{3}\phi}{2} \right).
\end{equation}
For the sake of consistence, we should also verify that the Klein-Gordon equation is satisfied. With metric (\ref{metric}), it is written as
\begin{equation}
\frac{1}{r'r^2} \left[ \partial_{\tau} \left(r'r^2\dot{\phi}\right) - \partial_R \left(\frac{r^2}{r'}\phi'\right) \right] + \partial_{\phi} V = 0.
\end{equation}
By doing $\phi'=0$, and using (\ref{solution2}), (\ref{condition'}) and (\ref{rho'}), it is straightforward to verify that it is reduced to an identity.

In the limit of small $\phi$ (when the collapse begins), (\ref{potential}) can be approximated by a free field potential
\begin{equation}
V(\phi) = \Lambda + \frac{m^2}{2} \phi^2 \quad \quad (\tau \rightarrow -\infty),
\end{equation}
with the scalar field mass given by $m = \sqrt{3\Lambda}/2$. Therefore, the blackhole mass (\ref{mass}) is
\begin{equation}
 M = \frac{2\sqrt{3}\pi}{m}. 
\end{equation}
The scalar field candidate for a unified description of the cosmological dark sector has a mass of the order of the Hubble parameter, $m \sim H_0 \sim 10^{-33}$ eV \cite{sigma1}. Therefore, it may only form a blackhole of cosmological size. For a scalar field of mass $m \sim 10^{-5}$ eV (the axion mass) we have $r_{g} \sim 1$ cm, corresponding to a blackhole of mass $M \sim 10^{25}$ kg (one Earth mass). This agrees with current theoretical estimates for the mass of dense axion stars, which ranges from $10^{-14} M_{\odot}$ to 1 solar mass \cite{dense}. For a scalar field of Planck mass, the horizon has the Planck length, and the blackhole has a Planck mass. A Higgs blackhole, if formed prior to the Higgs decay, would weight a megatonne and evaporate in a century.

\section*{Acknowledgements}

The authors are thankful to A. Saa for valuable criticisms and suggestions, to G. Dotti, M. M. Guzzo, O. L. G. Peres and M. G. Rodrigues for useful discussions, and to CNPq for financial support. SC is grateful to P. C. de Holanda and F. Sobreira for the hospitality at UNICAMP.

\end{document}